# Alternative Stacking Sequences in Hexagonal Boron Nitride

S. Matt Gilbert[1,2,3], Thang Pham[1,2,3,4], Mehmet Dogan[1,2], Sehoon Oh[1,2], Brian Shevitski[1,2,3,5], Gabe Schumm[1,2,3], Stanley Liu[1,2,3], Peter Ercius[5], Shaul Aloni[5], Marvin L. Cohen[1,2], and Alex Zettl[1,2,3,*]

[1] Department of Physics, University of California at Berkeley, Berkeley, CA 94720, USA
[2] Materials Sciences Division, Lawrence Berkeley National Laboratory, Berkeley, CA 94720, USA
[3] Kavli Energy NanoScience Institute at the University of California, Berkeley and the Lawrence Berkeley National Laboratory, Berkeley, CA 94720, USA
[4] Department of Materials Science and Engineering, University of California at Berkeley, Berkeley, CA 94720, USA
[5] Molecular Foundry, Lawrence Berkeley National Laboratory, Berkeley, CA 94720, USA
*Corresponding author: azettl@berkeley.edu

Abstract:

The relative orientation of successive sheets, i.e. the stacking sequence, in layered two-dimensional materials is central to the electronic, thermal, and mechanical properties of the material. Often different stacking sequences have comparable cohesive energy, leading to alternative stable crystal structures. Here we theoretically and experimentally explore different stacking sequences in the van der Waals bonded material hexagonal boron nitride (h-BN). We examine the total energy, electronic bandgap, and dielectric response tensor for five distinct high symmetry stacking sequences for both bulk and bilayer forms of h-BN. Two sequences, the generally assumed AA' sequence and the relatively unknown (for h-BN) AB (Bernal) sequence, are predicted to have comparably low energy. We present a scalable modified chemical vapor deposition method that produces large flakes of virtually pure AB stacked h-BN; this new material complements the generally available AA' stacked h-BN.

**Introduction**

In recent years there has been a dramatic resurgence in interest in van der Waals bonded layered materials, including graphite, boron-nitride, and transition metal dichalcogenides.[1,2] These materials display strong intraplane (typically covalent) bonding and weak van der Waals interplane bonding, which facilitates exfoliation into mono-layer or few-layer forms, and further allows custom stack-ups or laminations of sheets with different chemical composition or crystallographic orientation.[1–8] Even for a material composed of identical sheets, the stacking order of the successive sheets or layers, which may be translationally and/or rotationally shifted, can profoundly influence the overall physical properties.[9–12] For example, for naturally occurring graphite the usual stacking sequence is Bernal (AB) stacking, but rhombohedral (ABC) stacking is also possible, which has a completely different electronic band structure.[13,14]

Hexagonal boron nitride (h-BN) is structurally very similar to graphite, with successively stacked (and van der Waals bonded) sheets of hexagonally arranged $sp^2$-bonded boron and nitrogen.[5] However, unlike graphite, h-BN is purely synthetic, with a wide electronic band gap (hence the nickname "white graphite").[15–17] Virtually all synthesis methods for h-BN lead to an AA' stacking sequence, where atoms in one layer all lie directly above atoms in the next layer.[4,18–20] Successive layers are rotated such that all nitrogens lie above borons, and all borons lie above nitrogens.[20] Potential alternative stacking sequences for h-BN are of great theoretical and experimental interest.

Here we explore these alternative h-BN stacking sequences. We employ Density Functional Theory (DFT) to determine the total energy, electronic band structure, and dielectric tensor elements for five different high-symmetry h-BN stackings. We find that Bernal (AB) stacked h-BN has a total energy comparable to, and indeed a bit lower than, that for conventional AA′ stacked h-BN. This suggests that Bernal stacked h-BN should be stable. We present a modified Low-Pressure Chemical Vapor Deposition (LP-CVD) method that reliably produces large flakes of few-layer Bernal stacked h-BN. This new form of h-BN is characterized by electron diffraction and atomic-resolution transmission electron microscopy measurements.

**h-BN Stacking Sequences**

Fig. 1 schematically shows five high-symmetry stacking sequence possibilities for h-BN. Boron and nitrogen atoms are indicated in gold and blue, respectively. In each schematic, only two superposed atomic sheets or layers are shown, and for visual clarity atoms in the top layer have been drawn slightly smaller than atoms in the bottom layer. Because there is no nomenclature consensus in the literature for all of these stacking sequences, we here adopt a naming convention where "prime" denotes a 60° rotation, and the letters A and B are used in the standard way as for graphite.

AA is the simplest stacking sequence; here the atoms in consecutive layers exactly align (B on B, and N on N). AA′ is obtained from AA by rotating every other layer by 60° and aligning the hexagons (B on N). AB is obtained from AA by shifting every other layer, as demonstrated by the red arrow in the upper right panel of the figure, which yields a structure where half of the B atoms align with half of the N atoms, and the remaining half of the atoms align with the centers of the hexagons in the neighboring layers. Shifting every second layer in AA′, as demonstrated by the red arrows in the lower right two panels of Fig. 1, yields AB1′ and AB2′. In AB1′ (AB2′), all the B (N) atoms are aligned, and all the N (B) atoms align with the centers of the hexagons in the neighboring layers. Applying to AA the lateral shift that we use to obtain AB2′ from AA′ yields an equivalent structure to AB; therefore, there are a total of five physically distinct structures that preserve the three-fold symmetry. We reiterate that AB stacking, also called Bernal stacking, is the stacking sequence most prevalent in naturally-occurring crystalline graphite,[21] while virtually all h-BN produced today and used in laboratories worldwide has the stacking sequence AA'. [4,15,18,20]

**Theoretical Results**

Using density functional theory (DFT), we explore the total cohesive energy, electronic band structure, and dielectric response tensor for the infinite crystal (bulk) and bilayer forms of h-BN, in the five physically distinct stacking sequences presented in Fig. 1. Similar calculations for total energy and band structure have been published previously by various other groups; unfortunately since the results are highly sensitive to the computational details such as the exchange-correlation functional and the van der Waals scheme, they yield inconsistencies.[22–27]

Table A presents the total energies of bulk and bilayer h-BN in the different stacking sequences, as computed in DFT. We take the AA′ stacking to be the reference and set its energy to zero. We also include the results from previous studies for comparison. We find AB to be the ground state (i.e. it has the lowest overall energy) for both bulk and bilayer BN. However, AA′ is quite close in energy (3 meV per unit cell higher for bulk). AB1′ is next closest in energy (8 meV per unit cell higher for bulk). This hierarchy is consistent with some reports in the literature[24,27] but at odds with others[22,25]. We also find that AA and AB2′, while close to each other in energy, are significantly higher in energy than the other three stackings. This can be understood by simple arguments such as those suggested in Ref. 11. Aligning negatively charged nitrogen atoms with large electron clouds causes a repulsive Pauli interaction between the layers, increasing the total energies and the interlayer distances to 3.64 Å for AA and 3.54 Å for AB2′ (from 3.33 Å for AA′, AB and AB1′). We note that these two high-energy stackings are also energetically unstable, i.e. they do not correspond to local energy minima in the configuration landscape and relax to one of the stable configurations when the atoms are slightly perturbed.

Fig. 2 shows our band structure results for h-BN bilayers (assumed suspended in vacuum). The zeros of the energy levels are set to the vacuum level for each structure so that there is a common reference between the plots. For additional insight we have projected the Kohn-Sham wavefunctions onto the atomic orbitals for each point along the bands we have computed. The resulting atomic orbital characteristics are designated by red (blue) coloring for boron (nitrogen) n=2 (2s & 2p) orbitals. We observe that the top of the valence band mostly consists of nitrogen n=2 orbitals, and the bottom of the conduction band mostly consists of boron n=2 orbitals, which verifies the ionic character of the intralayer bonds in h-BN, and is consistent with a large-gap insulator.

We summarize the band gaps we have obtained for h-BN in its bulk and bilayer forms in Table B. Due to the underestimation of band gaps in DFT, the values tabulated are expected to be much lower than experimental measurements, which are in the 4 to 7 eV range. [22] However, the relative values of the computed gaps can be expected to be mostly accurate. For instance, in Refs. 28 and 24, the bandgaps are computed with and without a GW correction, respectively, and it is found that the GW correction adds about 1.6 eV to the value of each gap. We also expect the shapes of the bands to be correctly predicted by DFT; therefore, the predictions as to whether a band gap is direct or indirect are reliable. We find that both AA′ and AB have indirect gaps. However, the AB structure allows an optical transition at the K point with an energy only 1% larger than the band gap. This is a key feature distinguishing the electronic structure of AB stacking from AA' stacking in h-BN.

Table C summarizes our calculations for the dielectric tensor of bilayer and bulk h-BN, for the five selected stacking sequences. Because of the high symmetry of h-BN, the dielectric tensor only has two distinct non-zero values: the in-plane $\varepsilon_{xx}=\varepsilon_{yy}$ (which we denote as $\varepsilon_{\parallel}$), and the out-of-plane $\varepsilon_{zz}$ (which we denote as $\varepsilon_{\perp}$). We compute the dielectric tensor in two frequency limits: $\varepsilon_{\infty}$ and $\varepsilon_0$, for high and low frequency, respectively. In the Born-Oppenheimer approximation, for high-frequency electric field response, the nuclei can be thought of as stationary, and the electronic response dominates. In this case, the dielectric response is described by $\varepsilon_{\infty}$, which we compute via the density functional perturbation theory. [29] For low-frequency electric-field response, $\varepsilon_0$, we compute the relaxed-ion dielectric tensor, following the methodology proposed by Ref. 30.

We find that the dielectric tensor values for AA′, AB and AB1′ are very close, and larger than the values for AA and AB2′ by 0.2 to 0.4. This simple trend can be understood as a result of the difference in the interlayer distance between these two groups of stackings. In the former group, which has smaller interlayer distances, the dielectric material simply occupies a larger portion of the space, leading to a larger dielectric tensor. Our results for the relaxed-ion tensors indicate that the response of the nuclei to a finite electric field does not significantly depend on the stacking, further emphasizing the weakness of the interlayer ionic interactions. We have also compared our results with Ref. 31 for the AA′ stacking, which, to our knowledge, is the only available calculation of these values in the literature. Our results are in good agreement for the in-plane dielectric constant values, and in reasonable agreement for the out-of-plane values. For the ε values in the bilayer case, we have carefully corrected for the existence of vacuum in our simulations by applying the combined capacitor rules, as described in Ref. 31. For the slab thickness we have used twice the value of the interlayer distance in each stacking.

**Motivation for Experiment**

The total energy calculation results presented in Table A suggest that of the 5 high symmetry stacking sequences of h-BN, AA and AB2' structures are likely highly unstable. The remaining three sequences, AA', AB, and AB1' are relatively close in energy, with AB representing the ground state. The original synthesis of h-BN by Balmain in 1842[32,33] yielded material with AA' stacking, and this situation continues today: virtually all h-BN produced by all synthetic routes has the AA' stacking sequence. This begs the question, is a synthesis route possible that allows access to the distinct AB Bernal ground state structure, or possibly the slightly higher energy AB1' configuration?

Apart from the properties discussed above, AB Bernal-stacked h-BN has been predicted to possess a tunable bandgap, along with other unique optical features.[34,35] Previous experimental studies have observed tantalizing small pockets of AB-stacked bilayer h-BN.[35–37] Therefore, we are motivated to develop a viable synthetic route for the reliable synthesis of AB-stacked h-BN, a new form of two-dimensional insulator.

**Experimental Results**

We employ Low-Pressure Chemical Vapor Deposition (LP-CVD) on a transition metal surface to select the stacking growth sequence of h-BN. We find that the growth can be tuned from conventional AA' growth to exclusively Bernal AB growth, on both Cu and Fe substrates. High-Resolution Transmission Electron Microscopy (HRTEM) and Selected Area Electron Diffraction (SAED) are used to reveal the Bernal-stacking of the resulting multilayer flakes.

Fig. S1 shows the furnace setup used for the synthesis. Similar to Ref. 17 and our previous work[38], we employ a two-zone heating approach in which the gaseous thermal decomposition products of solid ammonia-borane (heated at 70-90 °C in region $T_1$ in Fig. S1) react to form h-BN on a transition metal catalyst (heated at at 1025 °C in region $T_2$ in Fig. S1). Our synthesis method for Bernal-stacked h-BN differs from previous ammonia-borane based LP-CVD growths of h-BN in its combination of relatively low precursor temperatures (75-85 °C) and high hydrogen gas flows (100-200 sccm). [17,39–41]

Figs. 3a-b show two representative Scanning Electron Microscope (SEM) images of few-layer h-BN flakes grown on Fe via the LP-CVD method (30 min, 100 sccm $H_2$). The concentric triangular shape and 40+ µm size of the flakes are representative of those grown over much of the entire metal substrate, with an average total substrate coverage by h-BN of approximately 50-90% depending on the growth conditions.

The stacking sequence of the h-BN flakes is examined by SAED and HRTEM. SAED has been demonstrated as an effective means to characterize the stacking of graphene.[42–44] By comparing the intensity of the first-order (($\pm100$), ($0\pm10$), and (±1∓10)) and second-order ((±2∓10), (0±10), and (±1∓20)) diffraction peaks, the stacking of the layers can be deduced.

Fig. 3c shows an SAED pattern of an h-BN flake originally grown via LP-CVD on an Fe substrate. The ratio of the intensities of the first-order <010> and second-order <110> diffraction peaks is ~0.3 (Fig. S2). We simulate the diffraction patterns of the stable bulk configurations of h-BN with AA′ and AB stackings using the Single Crystal software package (as shown in Fig. S2) and compare with our measured values. Although ABC-stacked rhombohedral BN does not belong to the h-BN class, we also simulate the diffraction patterns for it to rule out this stacking in our material.[45,46] The experimental first-order to second-order ratio of ~0.3 is consistent with the simulated ratio of 0.28 for AB stacked h-BN while the simulation of AA′ yields a ratio of 1.1 and ABC yields a ratio of 0. These ratios of intensities are comparable to results in graphene.[44] This result indicates that the as grown h-BN is either AB, AB1′, or AB2′ stacked, as they possess similar intensity ratios.

To differentiate between these stackings, we perform HRTEM on a bilayer region of material. Figs. 3d-e show aberration corrected HRTEM focal series reconstruction acquired at 80 kV (described in the methods section of the supplement) of two nearby mixed monolayer/bilayer regions of the multilayer h-BN as cropped from the full image shown in Fig. S3. The red dashed triangle denotes the boundary between the bilayer and monolayer/vacuum regions that were formed due to electron irradiation in the TEM.

The atomic positions in each region are deduced from Fig. 3d-e and are sketched in Fig. 3f. Based on the atomic positions as sketched, the two layers are AB stacked (see Fig. 1). We deduce the atomic positions shown in Fig. 3f by assuming that all edge-atoms of the triangular holes are N as is well-reported in the literature at ambient conditions. [20,38,47–50] We thereby determine the position of all other atoms in both layers of the bilayer area accordingly, as illustrated by the ball-and-stick model in Fig. 3f.

We note that the monolayer regions in Figs. 3d,e are two different layers. The regions shown in Fig 3d and 3e were close enough to be captured in the same HR-TEM (Fig. S3) image allowing us to directly correlate their structures. The atoms of the monolayer in Fig. 3d, which align with the referenced yellow line connecting bright contrast spots running through the entire bilayer area (the bright contrast spots are attributed to the overlapping of two atoms in the top and bottom layers), are all B. On the other hand, the atoms of the monolayer, which also align with the referenced yellow line as indicated in Fig. 3e, are all N. As the result, these monolayer regions reside in different layers. More importantly, one layer translates with respect to the other by a single atomic bond length (1.4 Å)[20] along the yellow reference line. This translation reflects AB (Bernal) stacking as shown in Fig. 1. This is the first conclusive observation to our knowledge of non-overlapping vacancies in two separate layers of a two-dimensional material.

In order to elucidate the growth mechanism for Bernal-stacked h-BN, we explore the role of increased hydrogen flow in our LP-CVD process. Fig. 4a shows SEM images from four h-BN samples grown on Cu with varying hydrogen gas flows (low 20 sccm, high 200 sccm) and growth time (20 min and 1 hour).

As shown in Fig. 4a(i-ii), low hydrogen flows, when combined with low precursor temperatures, lead to monolayer h-BN domains (Fig. 4a(i)) that merge into continuous films with additional growth time (Fig. 4a(ii)). Hydrogen gas is known to etch the edges of h-BN flakes at high temperatures. [39,51] Therefore, by utilizing a high hydrogen flow and low precursor temperature as in Fig. 4a(iii-iv), the edges of the h-BN flakes are etched, and the domains are prevented from merging into films. For short growth times (20 min), the triangular h-BN domains grown with high hydrogen flow (Fig. 4a(iii)) are smaller in comparison to those with low hydrogen flow (Fig. 4a(i)), but these smaller crystals have a higher incidence of ad-layers (white triangles within the larger, darker h-BN monolayer triangle). As the growth time is increased, the h-BN crystals grown with high-hydrogen-flow do not significantly increase in size, but the size and coverage of multilayer triangles increases as shown in Figure 2(c(iv)).

Based on these observations, we propose an under-layer growth model for Bernal-stacked h-BN, as sketched in Fig. 4b. This type of mechanism has been previously described for multilayer graphene and h-BN. [52–55] By maintaining low flows of precursor gas and high flows of hydrogen, our LP-CVD process can continuously etch the edges of the h-BN domains. This allows the gaseous precursor to access the transition metal catalyst for an extended period of time. The precursor gas can thereby adsorb onto or dissolve into the metal and form ad-layers below the first layer.

It has been previously reported that an h-BN layer grown on metal foil by CVD will have its orientation uniquely determined by the local configuration of the catalyst at the nucleation site of the h-BN.[56] Since each layer of h-BN grows directly on the catalyst in our process, its rotation can be controlled by the metal. Generally, we find that the layers of our Bernal-stacked h-BN grow as aligned triangles with a common centroid; this suggests the layers all form at a common nucleation point, forcing each layer to align to the metal. This results in aligned layers (which we have shown has a ground state of AB stacking) rather than anti-aligned (which has a ground state of AA').

We therefore conclude that our relatively high hydrogen gas flow and low precursor temperatures allow successive h-BN layers to form below the existing layers where they align to the metal resulting in Bernal-stacked h-BN. We note that this growth mechanism suggests that we could grow heterostructures of different materials with aligned stacking or single crystals with arbitrary twist angles by similar processes. By changing the precursor during the growth, the composition of the under-layer could be changed. By inducing nucleation at the grain boundaries of a polycrystalline metals, the orientation of the under-layer could be tuned.

**Acknowledgments**

This research was supported by the Director, Office of Science, Office of Basic Energy Sciences, Materials Sciences and Engineering Division, of the U.S. Department of Energy under Contract No. DE-AC02-05-CH11231, primarily within the sp2-Bonded Materials Program (KC-2207) which provided for growth of the h-BN and theoretical analysis; and in part by the van der Waals Heterostructures program (KCWF16) which provided for diffraction analysis. This work was additionally supported by the National Science Foundation under Grant # DMR-1206512 which provided for conventional TEM imaging; under Grant #1542741 which provided for construction of the LP-CVD synthesis system; and Grant #DMR 1508412 which provided for theoretical calculations. High-resolution TEM imaging at the Molecular Foundry was supported by the Office of Science, Office of Basic Energy Sciences, of the US Department of Energy under Contract No. DE-AC02-05CH11231. SMG acknowledges support from a Kavli NanoEnergy Sciences Institute Fellowship and an NSF Graduate Fellowship.

|  | Energy (meV per cell) | Method | AA | AA′ | AB | AB1′ | AB2′ |
|---|---|---|---|---|---|---|---|
| **bulk** | Present | GGA+TS vdW | 48 | ≡ 0 | -3 | 5 | 42 |
|  | Liu (2003)[23] | LDA | 56 | 0 | 0 | 8 | 48 |
|  | Ooi (2006)[22] | LDA | 51 | 0 | 6 | 25 | 48 |
|  | Hu (2011)[24] | LDA | 52 | 0 | -2 | 6 | 47 |
|  | Constantinescu (2013)[25] | LMP2 | 80 | 0 | 3 | 25 | 82 |
| **bilayer** | Present | GGA+TS vdW | 21 | ≡ 0 | -2 | 2 | 18 |
|  | Riberio (2011)[26] | GGA | 13 | 0 | 0 | 3 | 10 |
|  | Constantinescu (2013)[25] | LMP2 | 40 | 0 | 0 | 9 | 33 |
|  | Fujimoto (2016)[27] | LDA | 24 | 0 | -2 | 2 | 22 |

**Table A:** Computed total energies of the five high-symmetry stackings of hexagonal BN in its bulk and bilayer forms. In the present work, we use density functional theory (DFT) with the Perdew-Burke-Ernzerhof generalized gradient approximation (PBE GGA)[37] and norm-conserving pseudopotentials. To include the interlayer van der Waals (vdW) interactions, we have used the Tkatchenko-Scheffler (TS) correction as described in the supplement. Energies are referenced to the commonly known AA' stacking, and are in meV per unit cell (two B and two N atoms).

|  | Band gap (eV) | Method | AA | AA′ | AB | AB1′ | AB2′ |
|---|---|---|---|---|---|---|---|
| **bulk** | Present | GGA+TS vdW | 3.7248 (d) | 4.2932 (i) | 4.3751 (i) | 3.6470 (i) | 3.6245 (i) |
|  | Liu (2003)[23] | LDA | 3.226 (i) | 4.027 (i) | 4.208 (i) | 3.395 (d) | 3.433 (i) |
|  | Ooi (2006)[22] | GGA | 3.12 (d) | 4.28 (i) | 4.39 (i) | 3.66 (i) | 3.23 (i) |
|  | Hu (2011)[24] | LDA | 3.352 (i) | 4.015 (i) | 4.202 (i) | 3.407 (i) | 3.369 (i) |
|  | Hu (2012)[28] | GW | 4.96 | 5.73 | 5.87 | 5.01 | 4.96 |

| | | | | | | |
|---|---|---|---|---|---|---|
| **bilayer** | Present | GGA+TS vdW | 4.1986 (d) | 4.5249 (i) | 4.5032 (i) | 4.1432 (d) | 4.2569 (i) |
| | Riberio (2011)[26] | GGA | 4.23 | 4.69 | 4.60 | 4.29 | 4.52 |
| | Fujimoto (2016)[27] | LDA | 4.05 (d) | 4.34 (i) | 4.36 (i) | 4.01 (d) | 4.08 (i) |

**Table B:** Electronic band gaps for hexagonal BN in its five high-symmetry stackings, listed for the bulk and bilayer cases. Indirect and direct band gaps are denoted by "(i)" and "(d)", respectively.

| | | **AA** | **AA′** | **AB** | **AB1′** | **AB2′** | **AA′** [Laturia2018][31] |
|---|---|---|---|---|---|---|---|
| $\varepsilon_{\parallel,\infty}$ | Bulk | 4.41 | 4.71 | 4.80 | 4.78 | 4.49 | 4.98 |
| | bilayer | 4.40 | 4.68 | 4.74 | 4.75 | 4.43 | 4.97 |
| $\varepsilon_{\parallel,0}$ | bulk | 6.58 | 6.88 | 6.96 | 6.95 | 6.53 | 6.93 |
| | bilayer | 6.41 | 6.87 | 6.95 | 6.95 | 6.45 | 6.86 |
| $\varepsilon_{\perp,\infty}$ | bulk | 2.28 | 2.64 | 2.66 | 2.61 | 2.30 | 3.03 |
| | bilayer | 2.27 | 2.61 | 2.66 | 2.60 | 2.29 | 2.91 |
| $\varepsilon_{\perp,0}$ | bulk | 2.80 | 3.17 | 3.15 | 3.02 | 2.59 | 3.76 |
| | bilayer | 2.54 | 3.07 | 3.12 | 3.00 | 2.57 | 3.44 |

**Table C:** The high frequency (clamped-ion) and low frequency (free-ion) dielectric tensor elements for the five high-symmetry stackings of hexagonal BN. The values are relative to vacuum permittivity ($\varepsilon$ vacuum=1).

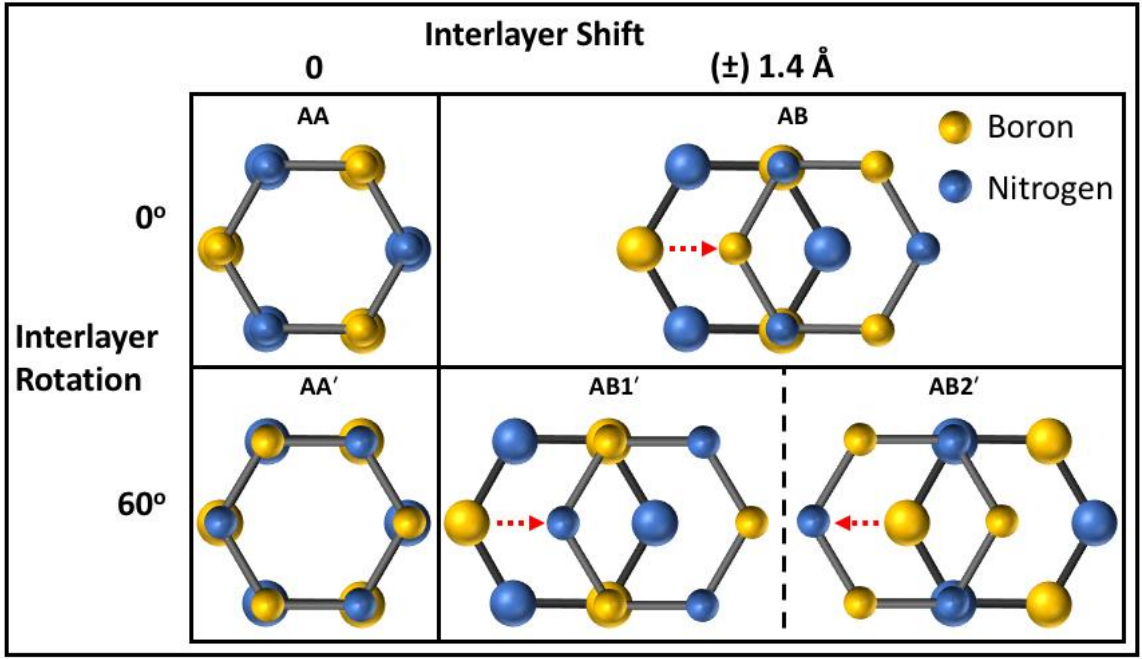

**Figure 1:** Schematics of the 5 high-symmetry stackings in bilayer h-BN as sorted by interlayer rotation and shift. In the top half of the figure, the rotationally aligned stacking configurations, AA and AB, are shown. AA is formed by stacking B to B and N to N in two aligned layers. AB is formed by translating one layer by a single bond length (1.4 Å)[20] to stack N to B as shown by the red arrow. In the bottom half of the figure, the rotationally anti-aligned stacking configurations, AA', AB1', and AB2', are shown. AA' is formed by stacking two anti-aligned layers B to N and N to B. AB1' is formed by translating one layer such that the layers stack B to B while AB2' is formed by translating one layer such that they stack N to N.

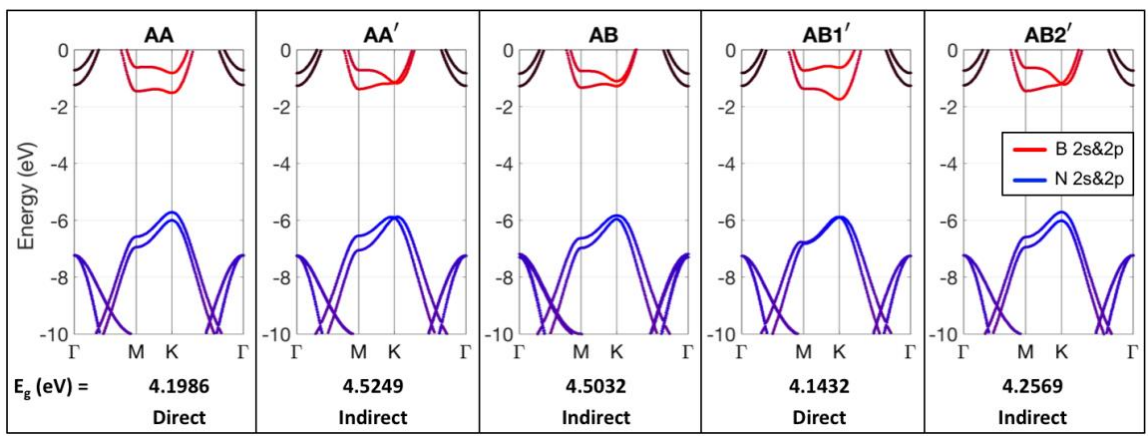

**Figure 2:** The electronic structure and bandgaps of the five physically distinct high-symmetry stackings of bilayer hexagonal boron nitride: AA, AA', AB, AB1', and AB2', from left to right. The red (blue) coloring of a band indicates the boron (nitrogen) 2s or 2p character of that band.

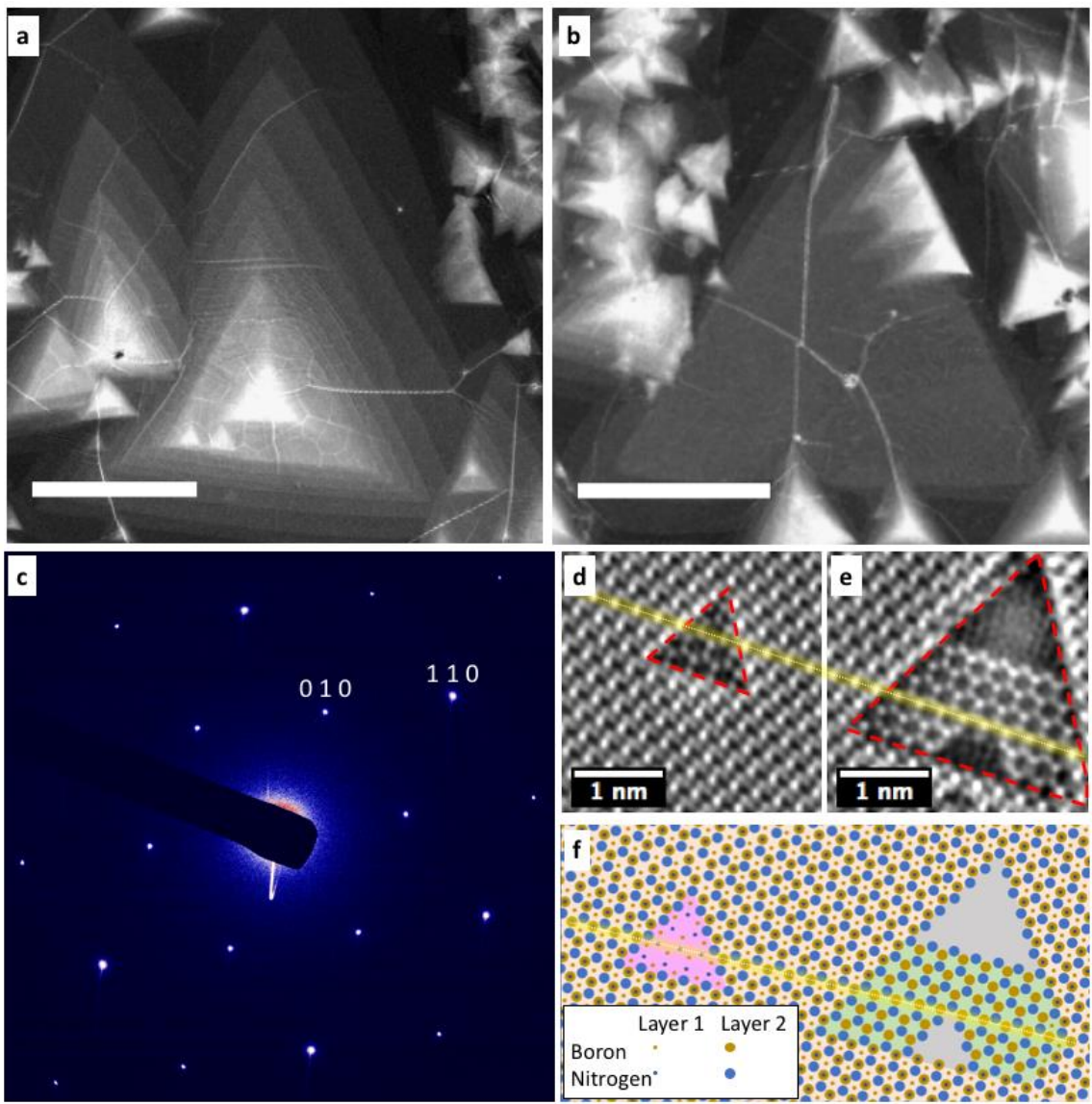

**Figure 3:** (a,b) SEM images of multilayer h-BN crystals as-grown on iron foil for 1 hour under 100 sccm $H_2$. The h-BN single crystals are bright contrasted triangles on a dark background of iron. Concentric triangles are additional layers stacked together. Brighter contrast indicates additional layers. (a) shows >10 separate layers stacked together with a large variation in width (2-30 µm) while (b) shows a more continuous region of multilayer h-BN over ~40 µm that is surrounded by smaller thicker h-BN triangles. (c) A representative SAED pattern for a thick region of multilayer LP-CVD h-BN. The ratio of the intensities of the first-order <010> and second order <110> peaks are compared to determine the stacking order of the h-BN. The ratio of ~0.3 as shown in Fig. S2 suggests that the stacking is AB, AB1', or AB2'. (d,e) Aberration corrected HR-TEM focal series reconstruction of two nearby mixed bilayer/monolayer regions of h-BN as cropped from the full image shown in Fig. S3. The red-dashed triangles denote the boundary between the monolayer/vacuum(inner) and the bilayer(outer) areas. (f) A schematic of the position of each atom in (d,e). The atomic positions are deduced as described in the text. The opaque yellow lines in (d,e,f) trace the line of stacked atoms in the bilayer region to highlight the alignment of the left and right monolayer regions. Scale bars are (a) 10 µm and (b) 15 µm.

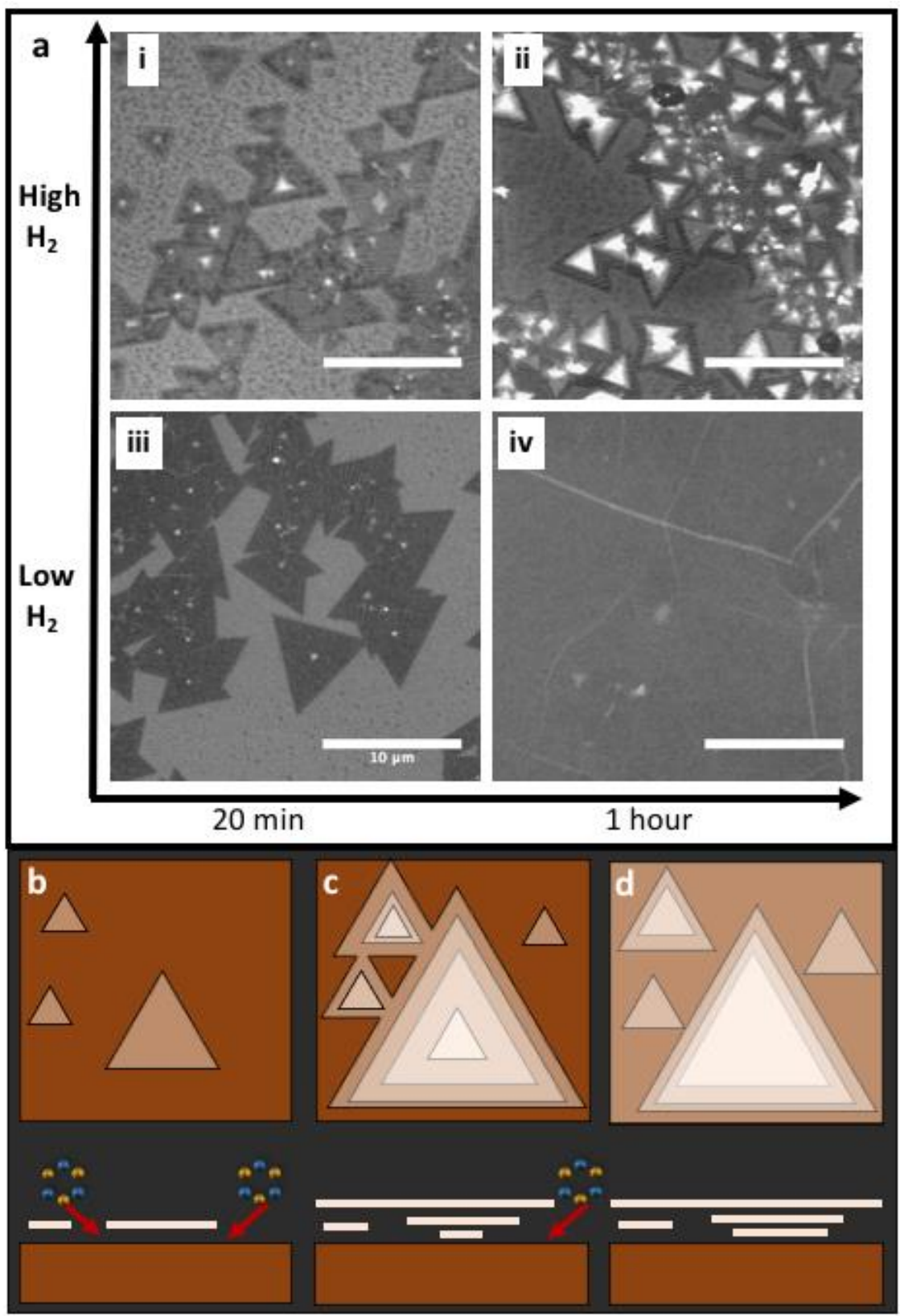

**Figure 4:** (a) SEM images of multilayer h-BN crystals grown on copper with varying times (20 min and 1 hour) and hydrogen flows (20 sccm and 200 sccm). (i) shows 3 µm triangular h-BN domains with small ad-layers (white) that are representative of a 20 minute growth with 200 sccm H2. (ii) shows that after 2 hours the ad-layers grow larger while the monolayer h-BN triangle remains a similar size under 200 sccm H2. (iii) shows 5 µm triangular h-BN domains that are representative of a 20 minute growth with 20 sccm H2. (iv) shows that after one hour with 20 sccm H2, the h-BN crystals merge to a full-coverage film with only minimal multi-layer coverage. (b-d) A cartoon illustrating the growth mechanism for Bernal-stacked h-BN from the top-view (above) and the side-view (below). (b) depicts the seeding and growth of h-BN monolayers. (c) shows the growth of the h-BN monolayers and the seeding of ad-layers underneath the original monolayer. (d) shows the h-BN film achieving full coverage thereby stopping the growth. Scale bars are 10 µm.